\documentclass[aps,pre,preprint,showpacs,superscriptaddress]{revtex4}
\usepackage{amsmath}
\usepackage{epsfig}

\newfont{\logo}{logo10}
\newcommand{\bear}{\begin{eqnarray}}
\newcommand{\eear}{\end{eqnarray}}
\newcommand{\bes}{\begin{subequations}}
\newcommand{\ees}{\end{subequations}}
\newcommand{\al}{\alpha}

\newcommand{\del}{\delta}
\newcommand{\s}{\sigma}

\begin{document}
\bibliographystyle{revtex}
\title{Shape changing (intensity redistribution) collisions of solitons in mixed coupled nonlinear Schr{\"o}dinger equations}

\author{T. Kanna\footnote{corresponding author, e-mail: Thambithurai.Kanna@u-bourgogne.fr}} 
\affiliation{Laboratoire de Physique de l'Universit{\'e} de Bourgogne, UMR CNRS No 5027,  \\
Av. A Savary, BP47 870, 21078 Dijon C{\'e}dex, France}
\author{M. Lakshmanan}
\affiliation{Centre for Nonlinear Dynamics, Department of Physics, Bharathidasan University, Tiruchirapalli-620 024, India\\} 
\author{P. Tchofo Dinda}
\affiliation{Laboratoire de Physique de l'Universit{\'e} de Bourgogne, UMR CNRS No 5027,  \\
Av. A Savary, BP47 870, 21078 Dijon C{\'e}dex, France}
\author{Nail Akhmediev}
\affiliation{ 
Optical Sciences Group, Research School of Physical
Sciences and Engineering, \\
The Australian National University, Canberra, ACT 0200, Australia}

\begin{abstract} 

A novel kind of shape changing (intensity redistribution) collision with potential application to signal amplification is identified in the integrable $N$-coupled 
nonlinear Schr{\"o}dinger (CNLS) equations with mixed signs of focusing and 
defocusing type nonlinearity coefficients. The corresponding soliton 
solutions for $N=2$ case are obtained  by using Hirota's bilinearization 
method. The distinguishing feature of the mixed sign CNLS equations is that the soliton solutions can both be singular and regular.  Although the general soliton solution admits singularities we present parametric 
 conditions for which non-singular soliton propagation can occur. 
The multisoliton solutions and a generalization of the 
results to multicomponent case with arbitrary $N$ are also presented. An appealing feature of soliton collision in the present case is that all the components of  a soliton can simultaneously  enhance their amplitudes, which can lead to new kind of amplification process without induced noise.
\end{abstract}
\pacs{02.30IK, 42.81Dp, 42.65Tg}
\maketitle
\section{Introduction}
It was suggested a long time ago that solitons could be used to carry data at very high bit rate in optical communication systems, because of their ability to overcome the dispersion limitation through a balance between the self-phase modulation and dispersion effects \cite{ad1}.
In fact soliton pulses are known to have many other desirable properties, such as their robustness against small changes in the pulse shape or amplitude around the exact soliton profile leads to treat such changes only as small perturbations on soliton propagation \cite{ad2,Mlb, Nailb}. Strictly speaking, the soliton properties can exit only in an ideal fiber. Indeed, in a standard telecommunication fiber, the propagation of light pulses gives rise to a host of perturbing effects which inhibit the desirable properties of solitons \cite{ad3}. One of the strongly perturbing effects that comes inevitably into play is the linear attenuation of light along the fiber (which is of the order of  0.2dB/km at carrier wavelength $1.55 \mu m$), which does not permit to keep a constant balance between the self-phase modulation and the group-velocity dispersion \cite{ad3}. Although the fundamental soliton propagation cannot be obtained in  standard fibers, pulse propagation over relatively long distances (and even transoceanic distances) can still be obtained through an appropriate combination of dispersion management and optical amplification (now mostly based on Er-doped fiber amplifiers and Raman amplifiers)  \cite{ad4, ad5, ad6}. 

All the existing amplification processes involve three major ingredients: The first one is a {\it pump wave}, which serves as a photon reservoir. The second one is an {\it amplification medium}, that is, a special material in which the pump wave is mixed with the signal. The third ingredient is a {\it physical mechanism} that can cause a transfer of photons from the pump to the signal. Only three types of physical mechanisms have been exploited so far in optical amplifiers, namely the {\it laser process} used in laser optical amplifiers (e.g. Er-doped fiber amplifiers, semi-conductor optical amplifiers) \cite{ad7}, the {\it stimulated Raman scattering} (used in Raman amplifiers) \cite{ad3, ad6} and {\it parametric wave mixing} (used in parametric amplifiers) \cite{ad3, ad6}. Such optical amplifiers do permit to fully compensate the fiber losses, but the amplification process is unavoidably accompanied by an undesirable effect of noise generation which is commonly referred to as the ''amplified spontaneous emission" (ASE) \cite{ad8, ad9, ad10}. Hence, one of the most important characteristic parameters of  the optical amplifiers developed so far is the so-called ''noise figure", which serves as a measure of the amount of noise generated during the amplification process \cite{ad11}. The ASE increases with the amplifier gain, and there exists an unavoidable amount of noise, known as the amplifier noise figure limit of 3 dB \cite{ad11, ad12, ad13}. The ASE is one of the major effects that severely degrades the transmission quality of ultra-short light pulses over long distances \cite{ad3, ad5, ad14}. To radically resolve the problem of ASE limitation in high-speed long-distance transmission systems, it is clear that the conceptual approach of optical amplification based on the three ingredients mentioned above needs to be partially or totally reformulated.

In the present work, we examine shape changing (intensity redistribution) collisions of vector solitons in mixed coupled 
nonlinear Schr{\"o}dinger (CNLS) equations, and report some results that suggest the possibility of constructing a novel approach of signal amplification. The novelty lies in viewing the collision process of solitons as a fundamental physical mechanism for transferring energy from the pump to the signal. The collision involves two vector solitons. One of the two solitons, say $S_1$, is chosen, to be the signal, while the other soliton ($S_2$) serves as the energy reservoir (pump wave). The major virtue of this type of collision-based amplification process is that it does not induce any noise, as it does not make use of any external amplification medium.

0n the other hand, the study of physical 
and mathematical aspects of CNLS
equations is of considerable current interest as these equations arise in diverse areas of science like nonlinear 
optics, optical communication, bio-physics, Bose-Einstein 
condensates, and plasma physics  \cite{Mahankov,Mlb, Nailb, Chaos, Scott}. The fundamental integrable $N$-CNLS system is given by the following set 
of equations
\begin{subequations}
\label{1}
\bear
  iq_{j,z}+q_{j,tt}+2 \mu\left(\sum_{l=1}^N\sigma_{l}|q_{l}|^2
  \right) q_{j} = 0, \, \;\; j=1,2,...,N, \label{1a}
\eear
where $q_j$, $j= 1,2, \dots, N$, is the complex amplitude of the $j$-th
component, the subscripts $z$ and $t$ denote the partial derivatives with
respect to normalized distance and retarded time,  respectively, $\mu$
represents the strength of nonlinearity ($\mu>0$)  and the coefficients
$\s_l$'s define the sign of the nonlinearity. System \eqref{1a} can be
classified into three classes as focusing,  defocusing and mixed types
depending on the signs of the nonlinearity  coefficients $\sigma_l$'s. The
focusing  case arises where all $\sigma_l$'s  are equal to $1$ and the
corresponding system admits bright soliton solutions   \cite{Manakov,
Radhakrishnan97, Akhmediev98, Kanna01,Kanna03, Park2000}.   These bright
solitons are found to undergo fascinating shape changing (intensity
redistribution) collisions~\cite{Radhakrishnan97, Kanna01, Kanna03} (for
other details see for example Refs.~\cite{at1,at2,at3}) and such collision 
properties are not observed in systems with defocusing nonlinearity  which
arises for all $\sigma_l$ $=$ $-1$ in \eqref{1a}. The latter system   possesses
either dark solitons in all the components or dark-bright  solitons which
undergo standard elastic collision  \cite{Radhakrishnan95,kiv, Park2000}.  Also
special analytic solutions for the focusing and defocusing types are given in
Refs. \cite{Hioea, Hioeb}. The third case arises for mixed  signs of
$\sigma_l$'s (that is, $+1$ or $-1$). For convenience,  we define $\sigma_l$'s
for this mixed case as
\bear
  \sigma_{l}&=&1\;\;\;\;\; \mbox{for} \;\;l=1, 2, \dots, n, 
\nonumber\\
  &=& -1 \;\;\;\mbox{for}\;\;l= n+1, n+2, \dots, N. \label{1b}
\eear
\end{subequations}
Here onwards we refer to Eq. \eqref{1} with the above choice of $\sigma_l$'s  
as {\em mixed} CNLS equations.

From a physical point of view, system \eqref{1} with $N=2$ corresponds to the
modified Hubbard model in one dimension \cite{Makhankov81a}.  Similar 
equation, for $N=2$, is observed in the context of electromagnetic pulse 
propagation in left handed materials \cite{Lazarides}.
The above set of equations \eqref{1}  is found to be completely integrable 
\cite{Makhankov81a, Makhankov81b, Zakharov} and the corresponding 
Lax pair was obtained in \cite{Makhankov81a}. In their pioneering 
works  Makhankov {\it et al.}~\cite{Makhankov81a,Makhankov81b} have 
shown that  Eq. \eqref{1}, for $N=2$, admits particular bright-bright,
bright-dark, dark-dark type one soliton solutions depending upon the 
asymptotic behaviour of the complex amplitudes $q_j$, $j=1,2$. Since then very
few works have appeared in the literature to analyse the problem further
\cite{Park2000, Radhakrishnan95, Hioe2002,  kanna2004, Tsoy} 
(for a detailed review of existing results one can refer to  
\cite{kanna2004}). Particularly, in a recent work \cite{kanna2004}, 
Kanna { \it et al.} have obtained stationary solutions of mixed CNLS 
equations with singularities by following an algebraic approach 
\cite{Akhmediev98, Nogami,  Akhmediev99}. It was observed 
that despite the points of singularities the solutions behave smoothly in finite region of the temporal domain.  Then the  natural question arises as to  whether
multisoliton solutions exhibiting regular behaviour over the entire space-time
regions exist and, if so, what is the nature of soliton interactions?
 
Being motivated by the above fundamental and intriguing aspects, in the present paper we 
perform a detailed study on the bright soliton collision dynamics arising 
in the mixed CNLS system. In particular, we point out that bright solitons
of regular type do exist, provided the soliton parameters satisfy certain
conditions and that the underlying solitons undergo novel shape changing/intensity
redistribution collisions.  The singular solutions turn out to be special cases (with
specific parametric choices) of the general soliton solutions. An important new feature which we identify in the collision process of regular solitons in the mixed CNLS case is that after collision a soliton can gain energy in all its components, while the opposite takes place in the other soliton.

This paper is organized as follows. Section 
II contains the details of Hirota's bilinearization procedure ~\cite{Hirota} 
for the CNLS equations to obtain soliton solutions. Though the solutions obtained 
in this paper admit both singular and non-singular behaviours, we call them 
as soliton solutions ascribing to their soliton nature in some specific 
region. In section III, we obtain the one and two soliton solutions. Section IV is devoted to
a detailed analysis of shape changing (intensity redistribution) collisions exhibited by these soliton 
solutions. The procedure to obtain one and two soliton solutions is extended 
to multisoliton solutions in section V. The results of 
two component case are generalized in a systematic way to the multicomponent 
case with arbitrary number of components  following the lines of Ref. 
\cite{Kanna03}. Final section is allotted for conclusion. In Appendix A we present the singular stationary three soliton solution  for mixed 3-CNLS equations. The multicomponent multisoliton solutions of mixed $N$-CNLS equations, for arbitrary $N$, is given in Appendix B. 

\section{Bilinearization of mixed CNLS equations}
The set of equations \eqref{1} has been shown to be completely integrable 
~\cite{Makhankov81a, Zakharov}, admitting certain  types of single soliton 
solutions  ~\cite{Makhankov81a,Makhankov81b}, for the $N=2$ case, as mentioned in the 
Introduction. Here we are concerned with bright-bright multisoliton solutions 
whose intensity profiles vanish asymptotically and with the nature of soliton
interactions.

Let us apply  the bilinearizing transformation \cite{Hirota}
\bear
q_j=\frac{g^{(j)}}{f},\;\; j=1,2,...,N, \label{2}
\eear
to Eq. (1) similar to the focusing case  $\s_l=1$, $l=1,2,..,N$ \cite{Kanna03}.  This results in the following set of bilinear equations,
\bes
\label{3}
\bear
(iD_z+D_t^2)g^{(j)}.f&=& 0,\;\; j=1,2,...,N, \label{3a}\\
D_t^2\left(f.f\right)&=& 2\mu\sum_{l=1}^N\sigma_l{g^{(l)}g^{(l)*}},  \label{3b}
\eear
where $\sigma_l$ is given by Eq. \eqref{1b}, $*$ denotes the complex conjugate, $g^{(j)}$'s  are
 complex functions, while
$f(z,t)$ is a real function and the Hirota's bilinear operators 
$D_z$ and $D_t$ are defined by 
\bear
D_z^nD_t^m(a.b)=\left(\frac{\partial}{\partial z}-\frac{\partial}{\partial z'}
\right)^n
\left(\frac{\partial}{\partial t}-\frac{\partial}{\partial t'}
\right)^ma(z,t)b(z',t')
\Big{\vert}_{(z=z', t=t')}.\label{4}
\eear
\ees
The above set of equations can be solved by introducing the following 
power series expansions for $g^{(j)}$'s and $f$:
\begin{subequations}
\label{5}
\begin{eqnarray}
g^{(j)} & = & \chi g_1^{(j)} + \chi^3 g_3^{(j)} +...,\;\; j=1,2,...,N,\\
f  &= &1 +\chi^2 f_2 +\chi^4 f_4+...,
\end{eqnarray}
\end{subequations}
where $\chi$ is the formal expansion parameter. The resulting set of 
equations, after collecting the terms with the same power in $\chi$, 
can be solved recursively to obtain the forms of $g^{(j)}$'s and $f$.
\section{Soliton solutions for N=2 case}
The mixed system  \eqref{1} with $N=2$ and $\s_1 = 1$, $\s_2=-1$ is of special physical interest. To start with, we consider this particular case. 
\subsection{One soliton solution}
In order to write down the one soliton solution we restrict the power series 
\eqref{5} to the lowest order
\bear
g^{(j)}=\chi g_1^{(j)}, \;j=1,2,\;\;\label{6}
f=1+\chi^2 f_2 .
\eear
\begin{figure}
	\centering
		\caption{Intensity plots of singular one soliton solution of Eq. \eqref{1} 
for $N=2$: (a) for the case $\vert \al_1^{(1)}\vert =\vert \al_1^{(2)}\vert$, 
(b) for the case $\vert \al_1^{(1)} \vert < \vert \al_1^{(2)} \vert$.}
	\label{fig:figure1}
\end{figure}
Then by solving the resulting set of linear partial differential 
equations recursively, one can write down the explicit 
one soliton solution as
\begin{subequations}
\bear
\left(
\begin{array}{c}
q_1\\
q_2 
\end{array}
\right) 
&= &
\left(
\begin{array}{c}
\alpha_1^{(1)}\\
\alpha_1^{(2)}
\end{array}
\right)\frac{e^{\eta_1}}{1+e^{\eta_1+\eta_1^*+R}} \label{po1} \\
&= &
\left(
\begin{array}{c}
A_1 \\
A_2
\end{array}
\right)
k_{1R}\,\mbox{sech}\,\left(\eta_{1R}+\frac{R}{2}\right)e^{i\eta_{1I}}, \label{7}
\eear
where 
\bear
\eta_1&=&k_1(t+ik_1z) = \eta_{1R}+i\eta_{1I}, A_j=\frac{\alpha_1^{(j)}}
{\left[\mu\left(\s_1|\al_1^{(1)}|^2+\s_2|\al_1^{(2)}|^2\right)\right]^{1/2}}, \;\;j=1,2, \\
e^R &=&\frac{\mu\left(\s_1|\al_1^{(1)}|^2+\s_2|\al_1^{(2)}|^2\right)}{(k_1+k_1^*)^2},\;\; \s_1=-\s_2=1.\label{7a}
\eear
Note that this one soliton solution is characterized by three arbitrary complex 
parameters $\alpha_1^{(1)}$,  $\alpha_1^{(2)}$, and $k_1 = k_{1R}+ik_{1I}$, where 
the suffices $R$ and $I$ represent the real and imaginary parts, respectively. The 
 quantities $k_{1R}A_1$ and $k_{1R}A_2$, give the amplitude of the soliton in components $q_1$ and $q_2$, respectively, subject to the condition 
\bear
\s_1|A_1|^2+\s_2|A_2|^2=\frac{1}{\mu}, \label{cons}
\eear
and the soliton velocity in each component 
is given by $2k_{1I}$. The position of the soliton is found to be 
\bear
\frac{R}{2k_{1R}}=\frac{1}{2k_{1R}}\mbox{ln}\left[
\frac{\mu\left(\s_1|\al_1^{(1)}|^2+\s_2|\al_1^{(2)}|^2\right)}{(k_1+k_1^*)^2}
\right]. \label{pos}
\eear
\end{subequations}
From Eq. \eqref{7}, it is clear that singular solutions start occurring
when $|\alpha_1^{(1)}|$ =  $|\alpha_1^{(2)}|$.  In this case,
one can easily observe from Eq. \eqref{7a} that the quantity 
$e^R$ becomes $0$, and one gets the solution
\begin{eqnarray}
\left(\begin{array}{c}q_1 \\q_2  \end{array}\right) = 
\left(\begin{array}{c}
\alpha_1^{(1)} \\
\alpha_1^{(2)}  
\end{array}\right)e^{\eta_1}
\end{eqnarray}
which is unbounded. Such an unbounded solution is depicted in Fig. \ref{fig:fig1}(a) for $k_1 = 1+i$,  $\alpha_1^{(1)} = \alpha_1^{(2)} = 1$, and $\mu = 1$.  

\begin{figure}
	\centering
		\caption{Intensity plots of regular one soliton solution of Eq. \eqref{1} for $ N=2$ case.}
	\label{fig:fig2}
\end{figure}

 When $|\alpha_1^{(1)}|$ $<$ $|\alpha_1^{(2)}|$, 
$e^R$ becomes negative (so $R$ becomes complex). In this case,
singularity occurs, whenever
\begin{subequations}
\begin{eqnarray}
1-|e^R| e^{2\eta_{1R}} &=& 0, 
\end{eqnarray}
{or }
\begin{eqnarray}
\eta_{1R} &=& \frac {1}{2} \mbox{ln} \left(\frac {1}{|e^R|}\right).
\end{eqnarray}
\end{subequations}
Again a singular solution in this case is plotted in Fig. \ref{fig:fig1}(b) for $k_1 = 1+i$,  $\alpha_1^{(1)} = 0.8$, $\alpha_1^{(2)} = 1$, and $\mu = 1$.   

However the bright soliton solution is always regular as long as the condition 
$|\alpha_1^{(1)}|$ $>$ $|\alpha_1^{(2)}|$ is valid in which case  $e^R$ is 
always real and positive, as the denominator $\left(1+e^{ \eta_{1}+\eta_{1}^*+R}\right)$ 
in Eq. \eqref{po1} is always positive definite (as $\eta_{1R}$ is real) for this choice. This regular one soliton solution is shown in Fig. \ref{fig:fig2} for $k_1 = 1+i$,  $\alpha_1^{(1)} = 1$, $\alpha_1^{(2)} = 0.2$, and $\mu = 1$. 

It is also interesting to note here that the polarization vector evolves 
in a hyperboloid defined by the surface $|A_1|^2-|A_2|^2=\frac{1}{\mu}$ 
\cite{Makhankov81a}, whereas in the Manakov case it is a sphere (that is
$|A_1|^2+|A_2|^2=\frac{1}{\mu}$)\cite{Kanna03} . This 
allows Eq. \eqref{1} to admit a rich variety of singular and non-singular 
solutions and makes significant difference in the collision scenario of 
bright solitons arising in the two systems as we will see in the 
following sections.

\subsection{Two soliton solution}
To obtain the two soliton solution  the power series expansion \eqref{5} is terminated at the higher order terms 
 \bes 
 \label{8}
\bear
g^{(j)}&=& \chi g_1^{(j)}+\chi^3 g_3^{(j)},\;\; j=1,2, \\
f &=& 1+\chi^2 f_2+\chi^4 f_4.
\eear
\ees
Then by solving the resultant linear partial differential equations 
recursively, we can write the explicit form of the solution as 
\begin{subequations}
\label{9}
\bear
q_j=\frac{\alpha_1^{(j)}e^{\eta_1}+\alpha_2^{(j)}e^{\eta_2}
+e^{\eta_1+\eta_1^*+\eta_2+\delta_{1j}}+e^{\eta_1+\eta_2+\eta_2^*
+\delta_{2j}}}
{D},\;\;\;j=1,2,
\eear 
where
\begin{eqnarray}
D =  1+e^{\eta_1+\eta_1^*+R_1}
+e^{\eta_1+\eta_2^*+\delta_0}
 +e^{\eta_1^*+\eta_2+\delta_0^*}+e^{\eta_2+\eta_2^*+R_2}
+e^{\eta_1+\eta_1^*+\eta_2+\eta_2^*+R_3}. \nonumber\\ \label{9b}
\end{eqnarray}
Various quantities found in Eq. \eqref{9}, are defined as below:
\begin{eqnarray}
\eta_i&=&k_i(t+ik_iz),\;\;
e^{\delta_0} = \frac{\kappa_{12}}{k_1+k_2^*},\;\;
e^{R_1} = \frac{\kappa_{11}}{k_1+k_1^*},\;\;\;\;
e^{R_2}=  \frac{\kappa_{22}}{k_2+k_2^*},\nonumber\\
e^{\delta_{1j}}&=&\frac{(k_1-k_2)(\alpha_1^{(j)}\kappa_{21}
-\alpha_2^{(j)}\kappa_{11})}{(k_1+k_1^*)(k_1^*+k_2)},\;\;
e^{\delta_{2j}}=
\frac{(k_2-k_1)(\alpha_2^{(j)}\kappa_{12}-\alpha_1^{(j)}\kappa_{22})}
{(k_2+k_2^*)(k_1+k_2^*)},\nonumber\\
e^{R_3}&=&  \frac{|k_1-k_2|^2}{(k_1+k_1^*)(k_2+k_2^*)|k_1+k_2^*|^2}
 (\kappa_{11}\kappa_{22}-\kappa_{12}\kappa_{21}), \label {rc1}
 \eear
\noindent and
\bear
\kappa_{ij}= \frac{\mu\left(\s_1\alpha_i^{(1)}\alpha_j^{(1)*}+\s_2\alpha_i^{(2)}\alpha_j^{(2)*}\right)}
{\left(k_i+k_j^*\right)},\;i,j=1,2, \label{10d}
\end{eqnarray}
\end{subequations}
where $\s_1 = 1$ and $\s_2=-1$. This solution is characterized by six 
arbitrary complex parameters $\alpha_1^{(1)}$, $\alpha_1^{(2)}$,  
$\alpha_2^{(1)}$,  $\alpha_2^{(2)}$, $k_1$, and $k_2$.  Note that the form
of the above two soliton solution remains the same as that of the Manakov case (where
$\sigma_1=+1$, $\sigma_2=+1$) \cite{Radhakrishnan97, Kanna03}, except for the crucial difference that in the 
expressions for the parameters $\kappa_{ij}$ in Eq. \eqref{10d} $\sigma_1=+1$ and
 $\sigma_2=-1$.

It can also be easily verified that the  singular stationary solution for 
the $N=2$ case given by Eq. (17) in Ref. \cite{kanna2004} can be obtained 
for the specific parametric choice
\bear
 \alpha_1^{(1)} = -e^{\eta_{10}},\;  
\alpha_2^{(2)}=  e^{\eta_{20}},\; \alpha_1^{(2)} = 0,  \;
\alpha_2^{(1)}= 0, \; k_{1I}=k_{2I}=0, \;\mu = 1,  \label{ss1}
\eear
where $\eta_{10}$ and $\eta_{20}$ are two 
arbitrary real parameters. For this choice of parameters, Eq. \eqref{9} reduces to the form
\begin{subequations}
\label{pcs1}
\begin{eqnarray}
q_1 & = & \frac{1}{\tilde{D}}\left(-e^{\eta_1}+\frac{(k_{1R}-k_{2R})e^{\eta_1+\eta_2+
\eta_2^*}}{4k_{2R}^2(k_{1R}+k_{2R})}\right),\\
q_2 & = &  \frac{1}{\tilde{D}}\left(e^{\eta_2}-\frac{(k_{1R}-k_{2R})e^{\eta_1+\eta_1^*+
\eta_2}}{4k_{1R}^2(k_{1R}+k_{2R})}\right),
\end{eqnarray}
where
\begin{eqnarray}
 \tilde{D}&=& 1+\left[\frac{e^{\eta_1+\eta_1^*}}{4k_{1R}^2}-
\frac{e^{\eta_2+\eta_2^*}}{4k_{2R}^2}\right]-\frac{(k_{1R}-k_{2R})^2
e^{\eta_1+\eta_1^*+\eta_2+\eta_2^*}}
{16k_{1R}^2k_{2R}^2(k_{1R}+k_{2R})^2},
\end{eqnarray}
and $\eta_j$ is redefined as
\begin{eqnarray}
\eta_j=k_{jR}(t+ik_{jR}z)+\eta_{j0},\;\;j=1,2,
\end{eqnarray}
\end{subequations}
where $\eta_{j0}$'s are arbitrary real parameters.
The above equation \eqref{pcs1} can be expressed in terms of hyperbolic functions as 
\bes
\label{pcs}
\bear
 q_1 &=& \frac{2 k_{1R}}{\hat{D}} \sqrt{\frac{k_{1R}+k_{2R}}{k_{1R}-k_{2R}}}\;\mbox{sinh}\left(k_{2R}t +\eta_{20}+\frac{1}{2}\mbox{ln}\left[\frac{k_{1R}-k_{2R}}{4k_{2R}^2(k_{1R}+k_{2R})} \right] \right)e^{ik_{1R}^2z}, \qquad \\
 q_2 &=& -\frac{2 k_{2R}}{\hat{D}} \sqrt{\frac{k_{1R}+k_{2R}}{k_{1R}-k_{2R}}}\;\mbox{sinh}\left(k_{1R}t +\eta_{10}+\frac{1}{2}\mbox{ln}\left[\frac{k_{1R}-k_{2R}}{4k_{1R}^2(k_{1R}+k_{2R})} \right] \right)e^{ik_{2R}^2z},\qquad 
 \eear 
 where 
 \bear
\hat{D} = -\mbox{sinh}\left(k_{1R}t+k_{2R}t+\eta_{10} +\eta_{20}+\mbox{ln}\left[\frac{k_{1R}-k_{2R}}{2k_{1R}k_{2R}(k_{1R}+k_{2R})} \right] \right) \nonumber\\
 +\left(\frac{k_{1R}+k_{2R}}{k_{1R}-k_{2R}}\right)\mbox{sinh}\left(k_{1R}t-k_{2R}t+\eta_{10} -\eta_{20}+\mbox{ln}\left[\frac{k_{2R}}{k_{1R}} \right]\right).
 \eear
\ees
One can check that Eq. (17) given in Ref. \cite{kanna2004} can be re-expressed in terms of hyperbolic functions in a  form similar to Eq. \eqref{pcs}. Figure \ref{fig:fig3} represents  the stationary singular two soliton solution at $z=0$ for $k_{1R} = 0.2$, $k_{2R} = -0.25 $, $\al_1^{(1)} = -\al_2^{(2)}= -1 $, $\al_1^{(2)} = \al_2^{(1)} = 0$, and $\mu = 1$.
\begin{figure}
	\centering
		\caption{Stationary singular two soliton solution  for $ N=2$ case.}
	\label{fig:fig3}
\end{figure}

Now from the expression \eqref{9} it can be observed that the  denominator 
can become zero for finite values of $z$ and $t$ leading to singular solutions.
However, in the case of the general two soliton solution (10), 
it is possible to make the denominator ($D$ in Eq. (10b)) to be non-zero for any value of $t$ and $z$ for suitable 
choice of $k_j$ and ${\alpha_j^{(l)}}$'s, $j, l=1,2$. In order to do so we rewrite the denominator $D$ (see Eq. \eqref{9b}) as 
\bes
\bear
\label{r2a}
D&=&  2 e^{\eta_{1R}+\eta_{2R}}\left\{ e^{(R_1+R_2)/2} \mbox{cosh}\left(\eta_{1R}-\eta_{2R}+(R_1-R_2)/2\right) \right. \nonumber\\
 &&\left.  +e^{\del_{0R}}\mbox{cos}\left(\eta_{1I}-\eta_{2I}+\del_{0I}\right) \right. \nonumber\\
&& \left.+e^{R_3/2}\mbox{cosh}\left(\eta_{1R}+\eta_{2R}+R_3/2\right) \right\},
\end{eqnarray}
where the suffices $R$ and $I$ denote the real and imaginary parts,  respectively. 
Then the solution is regular if the above expression is positive for all values
of $z$ and $t$.  For this purpose, a definite set of criteria can be identified
as follows.  As in the case of one soliton solution in Sec. IIIA, if we choose the
parameters $\alpha_i^{(j)}, i,j =1,2$, such that $|\alpha_i^{(1)}|^2 > 
|\alpha_i^{(2)}|^2, i=1,2$, $k_{1R} > 0$ and $k_{2R} > 0$ then
\bear
{\kappa}_{11} > 0,\;\;{\kappa}_{22} > 0.\label{r2b}
\eear
 Correspondingly, from Eqs. \eqref{rc1} we note that $e^{R_1} > 0$
and $e^{R_2} > 0$, so that
$e^{R_1+R_2} > 0$. Then, $e^{(R_1+R_2)/2} 
\mbox{cosh}\left(\eta_{1R}-\eta_{2R}+(R_1-R_2)/2\right) > 0$. 
There is also the other possibility 
$ {\kappa}_{11}< 0,\;\; {\kappa}_{22} < 0$. 
But it will not lead to regular solution as in this case $e ^{R_1}$ and $e ^{R_2}$ become negative thereby making $R_1$ and $R_2$ complex.

The term $e^{R_3/2}$ becomes greater than zero if 
\bear
 {\kappa}_{11}{\kappa}_{22} - |{\kappa}_{12}|^2 > 0. \label{r2c}
\eear
Then for this choice $e^{R_3/2}\mbox{cosh}\left(\eta_{1R}+\eta_{2R}+R_3/2\right)$ 
is always greater than zero. 

However, the term $ \mbox{cos}\left(\eta_{1I}-\eta_{2I}+\del_{0I}\right)$ oscillates between $-1$ and $1$. So in order that the middle term does not compensate the other two terms at any point in space/time resulting in $D$ being equal to zero, we should have 
 \bear
e^{(R_1+R_2)/2}+ e^{R_3/2} > e^{\del_{0R}}. \label{r2e}
\eear
Consequently using the expressions \eqref{rc1} in \eqref{r2e} one may deduce the condition 
\bear
\frac{1}{2}\sqrt{\frac{\kappa_{11}\kappa_{22}}{k_{1R}k_{2R}}}+ \frac{|k_1-k_2|}
{2|k_1+k_2^*|}\sqrt{\frac{{\kappa}_{11}{\kappa}_{22} - |{\kappa}_{12}|^2 }{{k_{1R}k_{2R}}}}
> \frac{|\kappa_{12}|}{|k_1+k_2^*|}.
 \label{r2f}
\eear
\ees
Note that the conditions \eqref{r2b} and \eqref{r2c} are necessary conditions
to obtain regular solution as their falsity will always result in singular
solution. Condition \eqref{r2f} is a sufficient one as its validity confirms
that the solution is always regular. We are unable to prove whether condition
\eqref{r2f} is also necessary or not due to the complicated form of the
function $D$ as a function of the variables $t$ and $z$ given by Eq. \eqref{9b}
or \eqref{r2a}. It appears that the latter can only be checked numerically for
given soliton parameter values. In terms of soliton parameters the conditions 
\eqref{r2b} and \eqref{r2c} read as 
\bes
\begin{eqnarray}
|\alpha_1^{(1)}|^2 - |\alpha_1^{(2)}|^2 &>& 0, \\
|\alpha_2^{(1)}|^2 - |\alpha_2^{(2)}|^2 &>& 0,
\end{eqnarray}
while \eqref{r2f} becomes
\begin{eqnarray}
\frac{(|\alpha_1^{(1)}|^2 - |\alpha_1^{(2)}|^2)(|\alpha_2^{(1)}|^2 - |\alpha_2^{(2)}|^2 )}
{|\alpha_1^{(1)}\alpha_2^{(1)*}-\alpha_1^{(2)}\alpha_2^{(2)*}|^2} &>& \frac{16k_{1R}^2k_{2R}^2}{(k_{1R}+k_{2R})^2+(k_{1I}-k_{2I})^2}. \nonumber\\
\end{eqnarray}
\ees
Thus the two soliton solution satisfying these conditions represent the interaction of two finite amplitude bright solitons with definite velocities and their collision behaviour is analysed in the follwing section.
\begin{figure}
\centering
\caption{Shape changing (intensity redistribution) collision of two solitons in the mixed CNLS system for $ N=2$ case.}
\label{fig:fig4}
\end{figure}

For illustrative purpose we consider the case $k_{1R}>0$, $k_{2R}>0$, $\mu = 1$, $\al_1^{(1)}= \mbox{\mbox{cosh}}(\theta_1)e^{i\phi_1}$, $\al_2^{(1)}= \mbox{\mbox{cosh}}(\theta_2)e^{i\phi_1}$, $\al_1^{(2)}= \mbox{sinh}(\theta_1)e^{i\phi_2}$, and $\al_2^{(2)}= \mbox{sinh}(\theta_2)e^{i\phi_2}$, for some arbitrary $\theta_1, \theta_2, \phi_1$, and $\phi_2$. Then, the conditions \eqref{r2b}, \eqref{r2c}, and \eqref{r2f} become 
 \bear
 {\kappa}_{11} &=& \frac{1}{2k_{1R}},\;{\kappa}_{22}= \frac{1}{2k_{2R}}, \;\;   \nonumber\\
 |k_1+k_2^*|^2 &- &4 k_{1R}k_{2R}\mbox{cosh}^2\left(\theta_{12}\right) > 0,
  \label{13} \nonumber\\
 \frac{1}{4k_{1R}k_{2R}}&+& \frac {|k_1-k_2|\sqrt{|k_1+k_2^*|^2-4k_{1R}k_{2R} \mbox{cosh}^2\left(\theta_{12}\right)}} 
{4k_{1R}k_{2R}|k_1+k_2^*|^2}  > 
\frac{\mbox{cosh}\left(\theta_{12}\right)}{|k_1+k_2^*|}, \label{13a}
 \eear
 where $\theta_{12} = \theta_{1}-\theta_{2}$.  A two soliton collision process corresponding to the condition \eqref{13a}  is shown in Fig. \ref{fig:fig4} for the parameter choice $k_1 = 1.0+i$,  $k_2 = 1.1-i$,  $\theta_1 =0.8$,
$\theta_2 =0.2 $, $\phi_1=1$ and $\phi_2=0.3$. This collision behaviour is analysed in detail in the following section.
\section{Shape changing (intensity redistribution) collisions of solitons}
Now it is of interest to understand the collision behaviour, shown in Fig. \ref{fig:fig4}, 
of the regular two soliton solution.  Figure  \ref{fig:fig4} shows the interaction
of two solitons $S_1$ and $S_2$ which are well separated before and after 
collision,  in the $q_1$ and $q_2$ components.  This figure shows that after collision,  the first soliton $S_1$ in the component $q_1$ gets enhanced
in its amplitude while the soliton $S_2$ is suppressed.  Interestingly, the same kind of changes 
are observed in the second component $q_2$ as well. This collision scenario is 
entirely different from the one observed in the Manakov system  where 
one soliton gets suppressed in one component and is enhanced in the other 
component with commensurate changes in the other soliton. 

On the other hand, conceptually, the collision scenario shown in Fig. \ref{fig:fig4} may be viewed as an amplification process in which the soliton $S_1$ represents a signal (or data carrier) while the soliton $S_2$ represents an energy reservoir (pump). The main virtue of this amplification process is that it does not require any external amplification medium and therefore the amplification of $S_1$ does not induce any noise.

The understanding of this fascinating collision process can be facilitated 
by making an asymptotic analysis of the two soliton solution as in the 
Manakov case \cite{Radhakrishnan97, Kanna03, Kanna01a}. We perform the analysis 
for the choice $k_{1R}, k_{2R}>0$ and $k_{1I}>k_{2I}$.  For any other choice
the analysis is similar. The study shows 
that due to collision, the amplitudes of the colliding solitons $S_1$ 
and $S_2$ 
change from $(A_1^{1-}k_{1R},A_2^{1-}k_{1R})$ and
 $(A_1^{2-}k_{2R},A_2^{2-}k_{2R})$ to  $(A_1^{1+}k_{1R},
A_2^{1+}k_{1R})$ and  $(A_1^{2+}k_{2R},A_2^{2+}k_{2R})$, respectively. 
Here  the superscripts in ${A_i^j}$'s denote the solitons (number(1,2)), the subscripts 
represent the components (number(1,2)) and '$\pm$' signs stand for '$z \rightarrow \pm \infty$'. They are defined as 
\begin{subequations}
\label{asy}
\bear
\left(
\begin{array}{c}
A_1^{1-}\\
A_2^{1-}
\end{array}
\right) & = & 
\left(
\begin{array}{c}
\alpha_1^{(1)} \\
\alpha_1^{(2)}
\end{array}
\right)
\frac{e^{-R_1/2}}{(k_1+k_1^*)}, \\
\left(
\begin{array}{c}
A_1^{2-}\\
A_2^{2-}
\end{array}
\right)&=&
\left(
\begin{array}{c}
 e^{\delta_{11}}\\
e^{\delta_{12}}
\end{array}
\right) \frac{e^{-(R_1+R_3)/2}}{(k_2+k_2^*)}, \\
\left(
\begin{array}{c}
A_1^{1+}\\
A_2^{1+}
\end{array}
\right)&=&
\left(
\begin{array}{c}
 e^{\delta_{21}}\\
e^{\delta_{22}}
\end{array}
\right)
\frac{e^{-(R_2+R_3)/2}}{(k_1+k_1^*)},\\
\left(
\begin{array}{c}
A_1^{2+}\\
A_2^{2+}
\end{array}
\right) &=&
\left(
\begin{array}{c}
\alpha_2^{(1)}\\
\alpha_2^{(2)}
\end{array}
\right) 
\frac{e^{-R_2/2}}{(k_2+k_2^*)}.
\eear
\ees
All the quantities in the above expressions  are given in
Eq. \eqref{9}
\cite{Radhakrishnan97, Kanna03, Kanna01a}.  
\begin{figure}
	\centering
	\caption{Elastic  collision of two solitons in the mixed CNLS system for  $N=2$ case.}
	\label{fig:fig5}
\end{figure}
The analysis reveals the fact that, for the non-singular two soliton solution, 
the colliding solitons change their amplitudes in each component according to the conservation equation
\bear
 |A_1^{j-}|^2 - |A_2^{j-}|^2 =  |A_1^{j+}|^2 - |A_2^{j+}|^2 = \frac{1}{\mu},\;\; j=1,2. \label{12}
 \eear 
This can be easily verified from the actual expressions given in Eq. \eqref{asy}.

This condition allows the given soliton to experience the same effect 
in each component during collision, which may find potential applications 
in some physical situations like noiseless amplification of a pulse.  
It can be easily observed from the conservation relation \eqref{12}
that each component of a given soliton
experiences the same kind of energy switching during collision process. 
The other soliton (say $ S_2$ ) experiences an opposite kind of energy 
switching due to the conservation law
\begin{eqnarray}
\int_{-\infty}^{\infty} |q_j|^2 dt =  \mbox{constant} , \;\;j=1,2,
\end{eqnarray}
as required from Eq. (1).

The asymptotic analysis also results in the following expression relating the intensities of solitons $S_1$ and $S_2$ in $q_1$ and $q_2$ components before and after interaction (see Eq. \eqref{asy}),
 \bear
|A_j^{l+}|^2=|T_j^l|^2|A_j^{l-}|^2,\;\;j,l=1,2, \label{13}
\eear
where the superscripts $l\pm$ represent 
the solitons designated as $S_1$ and $S_2$ at $z\rightarrow \pm\infty$.
The transition intensities are defined as
\bes
\label{14}
\begin{eqnarray}
|T_j^{1}|^2&=&\frac{|1-\lambda_2(\al_2^{(j)}/\al_1^{(j)})|^2}{
|1-\lambda_1\lambda_2|}, \\
|T_j^{2}|^2&=&\frac{|1-\lambda_1\lambda_2|}{
|1-\lambda_1(\al_1^{(j)}/\al_2^{(j)})|^2},\;\;\;
j=1,2,\\
\lambda_1&=&\frac{\kappa_{21}}{\kappa_{11}},\;\;\;
\;\;\; \lambda_2=\frac{\kappa_{12}}{\kappa_{22}}.
\end{eqnarray}
\ees
In fact, this way of energy (amplitude) redistribution can also be expressed in terms of linear fractional transformations (LFTs) as in the CNLS system with focusing nonlinearities \cite{Steg1,Steg2, Kanna03}. For example, one can identify from the asymptotic expressions \eqref{asy} that the state of $S_1$ after interaction (say $\rho_{1,2}^{1+} = \frac{A_1^{1+}}{A_2^{1+}}$) is related to its state before interaction (say $\rho_{1,2}^{1-} = \frac{A_1^{1-}}{A_2^{1-}}$)  through the following LFT,
\bes
\bear
\rho_{1,2}^{1+}=\frac{A_1^{1+}}{A_2^{1+}}=\frac{C_{11}^{(1)}\rho_{1,2}^{1-}
+C_{12}^{(1)}}
{C_{21}^{(1)}\rho_{1,2}^{1-}+C_{22}^{(1)}},\label{lft}
\eear
where
\bear
C_{11}^{(1)} &=&
\alpha_2^{(1)}\alpha_2^{(1)*}(k_2-k_1)+\alpha_2^{(2)}
\alpha_2^{(2)*}(k_1+k_2^*), \nonumber\\
C_{12}^{(1)} &=& -\alpha_2^{(1)}\alpha_2^{(2)*}(k_2+k_2^*), \nonumber\\ 
C_{21}^{(1)}& = &\alpha_2^{(2)}\alpha_2^{(1)*}(k_2+k_2^*), \nonumber\\
C_{22}^{(1)} &=& \alpha_2^{(2)}\alpha_2^{(2)*}(k_1-k_2)-\alpha_2^{(1)}\alpha_2^{(1)*}(k_1+k_2^*). 
\eear
\ees
A similar expression can be obtained for soliton $S_2$ also. The analysis of such state transformations preserving the difference of intensities among the components, during collision, in the context of optical computing and their advantage in constructing logic gates is kept for future study.

For the standard elastic collision  property ascribed to the scalar solitons 
to occur here we need the magnitudes of the transition intensities to be unity which is possible for 
the specific choice  
\bear
\frac{\al_1^{(1)}}{\al_2^{(1)}}=\frac{\al_1^{(2)}}{\al_2^{(2)}}. \label{15}
\eear
As an example in Fig. \ref{fig:fig5} we present the elastic collision for 
$\theta_1=\theta_2=0.2$, $\phi_1 = \phi_2 = 0.3$ (see Eq. \eqref{13a}), with $k_j$'s unaltered, 
$j=1,2$, (Note that this choice satisfies  the above condition \eqref{15}).
For all other values of $\alpha_i^{(j)}$'s, the soliton energies get exchanged
between the solitons in both the components as in Fig. \ref{fig:fig4}.  

The other quantities characterizing this collision process, along with 
this energy redistribution, are the amplitude dependent phase shifts 
and change in relative separation distances.  Their explicit forms can be
obtained as in the case of the Manakov model \cite{Radhakrishnan97, Kanna03}. Explicit expressions for 
the phase shifts $\Phi_1$ and $\Phi_2$ of solitons $S_1$ and $S_2$, 
respectively, during the collision are obtained from the asymptotic 
analysis as
\begin{eqnarray}
\Phi_{1} &=& -\Phi_{2} =\frac{\left(R_3-R_1-R_2\right)}{2}, \label{16}
\end{eqnarray}
where $R_1$, $R_2$, and $R_3$ are defined in Eq. \eqref{9}. 

Then, the change in relative separation distance between the solitons can be expressed as 
\begin{equation}
\Delta t_{12}=t_{12}^--t_{12}^+ =\frac{(k_{1R}+k_{2R})}{k_{1R}k_{2R}} \Phi_1, \label{17}
\end{equation}
where $t_{12}^{\pm}$ $=$ the position of $S_2$ (at $z \rightarrow \pm\infty$) minus position of
$S_1$  (at $z \rightarrow \pm \infty)$ .

\section{Generalization of the results to multisoliton solutions and multicomponent case}
Having discussed the nature of two soliton collision in the two component case ($N=2$), we now wish to study multisoliton collisions for the $N=2$ as well as $N >2$ cases. For this purpose, we will consider first the three soliton collision scenario for the $N=2$ case and then extend the analysis to more general cases.
\subsection{Multisoliton solutions}
It is straightforward to extend the  bilinearization procedure of 
obtaining one and two soliton solutions to multisoliton solutions as was done in Ref. \cite{Kanna03} for the integrable CNLS equations with focusing  nonlinearity coefficients. Below, we present  the form of the three soliton solution for the mixed CNLS equations \eqref{1} as
\begin{subequations}
\label{18}
\bear
q_j&=& \frac{\alpha_1^{(j)}e^{\eta_1}+\alpha_2^{(j)}e^{\eta_2}+\alpha_3^{(j)}
e^{\eta_3}
+e^{\eta_1+\eta_1^*+\eta_2+\delta_{1j}}+e^{\eta_1+\eta_1^*+\eta_3+\delta_{2j}}
+e^{\eta_2+\eta_2^*+\eta_1+\delta_{3j}}}
{D}\nonumber\\
&&+\frac{e^{\eta_2+\eta_2^*+\eta_3+\delta_{4j}}
+e^{\eta_3+\eta_3^*+\eta_1+\delta_{5j}}+e^{\eta_3+\eta_3^*+\eta_2+\delta_{6j}}
+e^{\eta_1^*+\eta_2+\eta_3+\delta_{7j}}
+e^{\eta_1+\eta_2^*+\eta_3+\delta_{8j}}
}{D} \nonumber\\
 &&+\frac{e^{\eta_1+\eta_2+\eta_3^*+\delta_{9j}}
+e^{\eta_1+\eta_1^*+\eta_2+\eta_2^*+\eta_3+\tau_{1j}}
+e^{\eta_1+\eta_1^*+\eta_3+\eta_3^*+\eta_2+\tau_{2j}}}{D}\nonumber\\
&&+\frac{
e^{\eta_2+\eta_2^*+\eta_3+\eta_3^*+\eta_1+\tau_{3j}}}{D},\;\;j=1,2,
\eear
where
\bear
D &=&1+e^{\eta_1+\eta_1^*+R_1}+e^{\eta_2+\eta_2^*+R_2}+e^{\eta_3+\eta_3^*+R_3}
+e^{\eta_1+\eta_2^*+\del_{10}}+e^{\eta_1^*+\eta_2+\del_{10}^*}\nonumber\\
&&+e^{\eta_1+\eta_3^*+\del_{20}}
+e^{\eta_1^*+\eta_3+\del_{20}^*}
+e^{\eta_2+\eta_3^*+\del_{30}}
+e^{\eta_2^*+\eta_3+\del_{30}^*}
+e^{\eta_1+\eta_1^*+\eta_2+\eta_2^*+R_4}\nonumber\\
&&+e^{\eta_1+\eta_1^*+\eta_3+\eta_3^*+R_5}
+e^{\eta_2+\eta_2^*+\eta_3+\eta_3^*+R_6}
+e^{\eta_1+\eta_1^*+\eta_2+\eta_3^*+\tau_{10}}
+e^{\eta_1+\eta_1^*+\eta_3+\eta_2^*+\tau_{10}^*}\nonumber\\
&&+e^{\eta_2+\eta_2^*+\eta_1+\eta_3^*+\tau_{20}}
+e^{\eta_2+\eta_2^*+\eta_1^*+\eta_3+\tau_{20}^*}
+e^{\eta_3+\eta_3^*+\eta_1+\eta_2^*+\tau_{30}}
+e^{\eta_3+\eta_3^*+\eta_1^*+\eta_2+\tau_{30}^*}\nonumber\\
&&+e^{\eta_1+\eta_1^*+\eta_2+\eta_2^*+\eta_3+\eta_3^*+R_7}. \label{18a}
\eear
\ees
Expressions for various quantities given in Eq. \eqref{18} 
have the following forms:
\bes
\bear
\eta_i&=&k_i(t+ik_iz), i=1,2,3,\\
e^{\delta_{1j}}&=&\frac{(k_1-k_2)(\al_1^{(j)}\kappa_{21}-\al_2^{(j)}\kappa_{11}
)}{(k_1+k_1^*)(k_1^*+k_2)},\;\;
e^{\delta_{2j}}=\frac{(k_1-k_3)(\al_1^{(j)}\kappa_{31}-\al_3^{(j)}\kappa_{11}
)}{(k_1+k_1^*)(k_1^*+k_3)},\nonumber\\
e^{\delta_{3j}}&=&\frac{(k_1-k_2)(\al_1^{(j)}\kappa_{22}-\al_2^{(j)}\kappa_{12}
)}{(k_1+k_2^*)(k_2+k_2^*)},\;\;
e^{\delta_{4j}}=\frac{(k_2-k_3)(\al_2^{(j)}\kappa_{32}-\al_3^{(j)}\kappa_{22}
)}{(k_2+k_2^*)(k_2^*+k_3)},\nonumber\\
e^{\delta_{5j}}&=&\frac{(k_1-k_3)(\al_1^{(j)}\kappa_{33}-\al_3^{(j)}\kappa_{13}
)}{(k_3+k_3^*)(k_3^*+k_1)},\;\;
e^{\delta_{6j}}=\frac{(k_2-k_3)(\al_2^{(j)}\kappa_{33}-\al_3^{(j)}\kappa_{23}
)}{(k_3^*+k_2)(k_3^*+k_3)},\nonumber\\
e^{\delta_{7j}}&=&\frac{(k_2-k_3)(\al_2^{(j)}\kappa_{31}-\al_3^{(j)}\kappa_{21}
)}{(k_1^*+k_2)(k_1^*+k_3)},\;\;
e^{\delta_{8j}}=\frac{(k_1-k_3)(\al_1^{(j)}\kappa_{32}-\al_3^{(j)}\kappa_{12}
)}{(k_1+k_2^*)(k_2^*+k_3)},\nonumber\\
e^{\delta_{9j}}&=&\frac{(k_1-k_2)(\al_1^{(j)}\kappa_{23}-\al_2^{(j)}\kappa_{13}
)}{(k_1+k_3^*)(k_2+k_3^*)},\nonumber\\
e^{\tau_{1j}}&=&\frac{(k_2-k_1)(k_3-k_1)(k_3-k_2)(k_2^*-k_1^*)}
{(k_1^*+k_1)(k_1^*+k_2)(k_1^*+k_3)(k_2^*+k_1)(k_2^*+k_2)(k_2^*+k_3)}\nonumber\\
&&\times
\left[\al_1^{(j)}(\kappa_{21}\kappa_{32}-\kappa_{22}\kappa_{31})
+\al_2^{(j)}(\kappa_{12}\kappa_{31}-\kappa_{32}\kappa_{11})
+\al_3^{(j)}(\kappa_{11}\kappa_{22}-\kappa_{12}\kappa_{21})
\right],\nonumber\\
e^{\tau_{2j}}&=&\frac{(k_2-k_1)(k_3-k_1)(k_3-k_2)(k_3^*-k_1^*)}
{(k_1^*+k_1)(k_1^*+k_2)(k_1^*+k_3)(k_3^*+k_1)(k_3^*+k_2)(k_3^*+k_3)}\nonumber\\
&&\times
\left[\al_1^{(j)}(\kappa_{33}\kappa_{21}-\kappa_{31}\kappa_{23})
+\al_2^{(j)}(\kappa_{31}\kappa_{13}-\kappa_{11}\kappa_{33})
+\al_3^{(j)}(\kappa_{23}\kappa_{11}-\kappa_{13}\kappa_{21})
\right],\nonumber\\
e^{\tau_{3j}}&=&\frac{(k_2-k_1)(k_3-k_1)(k_3-k_2)(k_3^*-k_2^*)}
{(k_2^*+k_1)(k_2^*+k_2)(k_2^*+k_3)(k_3^*+k_1)(k_3^*+k_2)(k_3^*+k_3)}\nonumber\\
&&\times
\left[\al_1^{(j)}(\kappa_{22}\kappa_{33}-\kappa_{23}\kappa_{32})
+\al_2^{(j)}(\kappa_{13}\kappa_{32}-\kappa_{33}\kappa_{12})
+\al_3^{(j)}(\kappa_{12}\kappa_{23}-\kappa_{22}\kappa_{13})
\right],\nonumber\\
e^{R_m}&=&\frac{\kappa_{mm}}{k_m+k_m^*}, \;\;m=1,2,3,\;\;
e^{\del_{10}}=\frac{\kappa_{12}}{k_1+k_2^*},\;\;
e^{\del_{20}}=\frac{\kappa_{13}}{k_1+k_3^*},\;\;
e^{\del_{30}}=\frac{\kappa_{23}}{k_2+k_3^*},\nonumber\\
e^{R_4}&=&\frac{(k_2-k_1)(k_2^*-k_1^*)}
{(k_1^*+k_1)(k_1^*+k_2)(k_1+k_2^*)(k_2^*+k_2)}
\left[\kappa_{11}\kappa_{22}-\kappa_{12}\kappa_{21}\right],\nonumber\\
e^{R_5}&=&\frac{(k_3-k_1)(k_3^*-k_1^*)}
{(k_1^*+k_1)(k_1^*+k_3)(k_3^*+k_1)(k_3^*+k_3)}
\left[\kappa_{33}\kappa_{11}-\kappa_{13}\kappa_{31}\right],\nonumber\\
e^{R_6}&=&\frac{(k_3-k_2)(k_3^*-k_2^*)}
{(k_2^*+k_2)(k_2^*+k_3)(k_3^*+k_2)(k_3+k_3^*)}
\left[\kappa_{22}\kappa_{33}-\kappa_{23}\kappa_{32}\right],\nonumber\\
e^{\tau_{10}}&=&\frac{(k_2-k_1)(k_3^*-k_1^*)}
{(k_1^*+k_1)(k_1^*+k_2)(k_3^*+k_1)(k_3^*+k_2)}
\left[\kappa_{11}\kappa_{23}-\kappa_{21}\kappa_{13}\right],\nonumber\\
e^{\tau_{20}}&=&\frac{(k_1-k_2)(k_3^*-k_2^*)}
{(k_2^*+k_1)(k_2^*+k_2)(k_3^*+k_1)(k_3^*+k_2)}
\left[\kappa_{22}\kappa_{13}-\kappa_{12}\kappa_{23}\right],\nonumber\\
e^{\tau_{30}}&=&\frac{(k_3-k_1)(k_3^*-k_2^*)}
{(k_2^*+k_1)(k_2^*+k_3)(k_3^*+k_1)(k_3^*+k_3)}
\left[\kappa_{33}\kappa_{12}-\kappa_{13}\kappa_{32}\right],
\eear
\bear
e^{R_7}&=& \frac{|k_1-k_2|^2|k_2-k_3|^2|k_3-k_1|^2}
{(k_1+k_1^*)(k_2+k_2^*)(k_3+k_3^*)|k_1+k_2^*|^2|k_2+k_3^*|^2|k_3+k_1^*|^2}
\nonumber\\
&&\times\left[(\kappa_{11}\kappa_{22}\kappa_{33}-
\kappa_{11}\kappa_{23}\kappa_{32})
+(\kappa_{12}\kappa_{23}\kappa_{31}-
\kappa_{12}\kappa_{21}\kappa_{33})\right .\nonumber\\
&&\left.+(\kappa_{21}\kappa_{13}\kappa_{32}-
\kappa_{22}\kappa_{13}\kappa_{31})\right],
\eear
and
\bear
\kappa_{ij}= \frac{\mu\sum_{l=1}^2 \sigma_l \alpha_i^{(l)}\alpha_j^{(l)*}} 
{\left(k_i+k_j^*\right)},\;i,j=1,2,3, \label{18b}
\eear
\end{subequations}
where $\s_1 = 1$ and $\s_2 = -1$. Here $\alpha_1^{(j)}$, $\alpha_2^{(j)}$ and
$\alpha_3^{(j)}$, $k_1$, $k_2$ and $k_3$, $j = 1, 2, 3$, are complex
parameters.

The solution \eqref{18} also features singular and non-singular behaviours, as
in the case of one and two soliton solutions depending upon the values of the
soliton parameters.  Though the denominator $D$ in the solution \eqref{18} is
cumbersome, possible  non-singular conditions can be obtained with some
effort.  Eq. \eqref{18a} can be rewritten as 
\bes
\bear
D&=&  2e^{\eta_{1R}+\eta_{2R}+\eta_{3R}}\left\{ e^{(R_1+R_6)/2} \mbox{cosh}\left(\eta_{1R}-\eta_{2R}-\eta_{3R}+(R_1-R_6)/2\right) \right. \nonumber\\
 &&\left.  
 +e^{(R_2+R_5)/2} \mbox{cosh}\left(\eta_{2R}-\eta_{1R}-\eta_{3R}+(R_2-R_5)/2\right)  
\right. \nonumber\\
&&\left.  +e^{(R_3+R_4)/2} \mbox{cosh}\left(\eta_{3R}-\eta_{1R}-\eta_{2R}+(R_3-R_4)/2\right)  \right. \nonumber\\
&&\left.  +2 e^{(\del_{10R}+\tau_{30R})/2}  \left(\mbox{cosh}(X_1)\mbox{cos}(Y_1)\mbox{cos}(Z_1) - \mbox{sinh}(X_1)\mbox{sin}(Y_1)\mbox{sin}(Z_1)\right)\right. \nonumber\\
&&\left.  +2e^{(\del_{20R}+\tau_{20R})/2} \left(\mbox{cosh}(X_2)\mbox{cos}(Y_2)\mbox{cos}(Z_2) - \mbox{sinh}(X_2)\mbox{sin}(Y_2)\mbox{sin}(Z_2)\right)\right. \nonumber\\
&&\left.  +2e^{(\del_{30R}+\tau_{10R})/2} \left(\mbox{cosh}(X_3)\mbox{cos}(Y_3)\mbox{cos}(Z_3) - \mbox{sinh}(X_3)\mbox{sin}(Y_3)\mbox{sin}(Z_3)\right)\right. \nonumber\\ &&\left.+e^{R_7/2} \mbox{cosh}\left(\eta_{1R}+\eta_{2R}+\eta_{3R}+R_7/2 \right)  \right\},
\end{eqnarray}
where 
\begin{eqnarray}
X_1 &=& -\eta_{3R}+\frac{(\del_{10R}-\tau_{30R})}{2}, \;\;\;
X_2 = -\eta_{2R}+\frac{(\del_{20R}-\tau_{20R})}{2}, \nonumber\\
X_3 &=& -\eta_{1R}+\frac{(\del_{30R}-\tau_{10R})}{2}, \;\;\;
Y_1 = \eta_{1I}-\eta_{2I}+\frac{(\del_{10I}+\tau_{30I})}{2}, \nonumber\\
Y_2 &=& \eta_{1I}-\eta_{3I}+\frac{(\del_{20I}+\tau_{20I})}{2}, \;\;\;
Y_3 = \eta_{2I}-\eta_{3I}+\frac{(\del_{30I}+\tau_{10I})}{2}, \nonumber\\
Z_1 &=& \frac{(\del_{10I}-\tau_{30I})}{2},\;\;\; Z_2 = \frac{(\del_{20I}-\tau_{20I})}{2}, \;\;\;
Z_3 = \frac{(\del_{30I}-\tau_{10I})}{2}.
\end{eqnarray}
\ees
Here the suffices $R$ and $I$ denote the real and imaginary parts, respectively.
\begin{figure}
\centering
		\caption{Shape changing (intensity redistribution) collision of three solitons in the mixed CNLS system for  $N=2$ case.}
	\label{fig:fig6}
\end{figure}
As in the case of two soliton solution here also we find the following conditions need to be satisfied for the solution to be regular:
\bes
\begin{eqnarray}
e^{R_i} > 0, \;\;i=1,2,...,7, \label{c1}\\
e^{(R_1+R_6)/2} ,  e^{(R_2+R_5)/2}, e^{(R_3+R_4)/2}, e^{R_7/2} &>& 4 \; \mbox{max}\left\{ e^{\del_{10R}+\tau_{30R}},e^{\del_{20R}+\tau_{20R}}, e^{\del_{30R}+\tau_{10R}}\right\}.\label{c2}\nonumber\\
\end{eqnarray}
\ees
Note that, the  conditions given in \eqref{c1} are necessary as the falsity of
any of them  always results in singular solution and the last condition
\eqref{c2} is sufficient to ensure that the given solution is regular. In fact
these conditions can also be expressed in terms of soliton parameters, but due
to their cumbersome nature we do not present them here. The appropriate choice
of parameters can be made by carefully looking at the explicit forms of 
$e^{R_i}, e^{\del_{j0}}$, and $e^{\tau_{j0}}, i = 1,...,7$, and $j=1,2,3$.

Such a non-singular solution representing the shape changing (intensity
redistribution)  collision of three solitons $S_1$, $S_2$, and $S_3$ in the two
components $q_1$ and $q_2$ is shown in Fig. \ref{fig:fig6} for the parameter
choice  $k_1 = 1+i$, $k_2 = 1.2-0.5i$, $k_3 = 1-i$, $\mu = 1$, $\alpha_1^{(1)}
=\mbox{cosh}(\theta_1)e^{i\phi_1}$,  $\alpha_2^{(1)} =\mbox{cosh}(\theta_2)
e^{i\phi_1}$, $\alpha_3^{(1)}  =\mbox{cosh}(\theta_3)e^{i\phi_1}$, 
$\alpha_1^{(2)} = \mbox{sinh}(\theta_1)e^{i\phi_2}$ , $\alpha_2^{(2)}
=\mbox{sinh}(\theta_2)e^{i\phi_2}$,  $\alpha_3^{(2)} =
\mbox{sinh}(\theta_3)e^{i\phi_2}$, where $\theta_1 = 0.8$, $\theta_2 = 0.4$,
$\theta_3 = 0.2$, $\phi_1 = 0.5$, and $\phi_2 = 1.0$.  From the figure we
observe that after collision solitons $S_1$ and $S_2$ are enhanced in their
intensities while there occurs suppression of intensity for soliton $S_3$ in
both the components $q_1$ and $q_2$. It can be verified that before and after
collision the conservation relation
\bear
 |A_1^{j-}|^2 - |A_2^{j-}|^2 = |A_1^{j+}|^2 - |A_2^{j+}|^2 = \frac{1}{\mu},\;\; j=1,2,3, \label{19}
\eear 
is satisfied, so that the difference of intensities of the solitons between
the  components $q_1$ and $q_2$ is preserved before and after the collision
process. The standard elastic collision can be regained if
$\alpha_1^{(1)}:\alpha_2^{(1)}:\alpha_3^{(1)}=\alpha_2^{(1)}:\alpha_2^{(2)}:\alpha_3^{(2)}.
$ Fig. \ref{fig:fig7} illustrates such an elastic collision for the choice
$\theta_1 = \theta_2 = \theta_3 = 0.4$, $\phi_1 = \phi_2 = 0.5$, with same
$k_j$'s , $j=1,2,3$, as in Fig. \ref{fig:fig6}.

\begin{figure}
	\centering
	\caption{Elastic collision of three solitons in the mixed CNLS system for  $N=2$ case.}
	\label{fig:fig7}
\end{figure}
In a similar manner the four soliton solution can be deduced from Eq.(A2) given in Ref. \cite{Kanna03} by redefining $\kappa_{ij}$ as in Eq. \eqref{18b} with $i,j$ running from $1$ to $4$. We do not present the explicit form of it here because of its cumbersome nature.
\subsection{Multicomponent case with N$>$2}
The next step is to generalize the above results for the $N=2$ case to arbitrary  $N$ with $N > 2$. To do this we follow the earlier work of two of  the authors (T.K. and M.L)\cite{Kanna03} on the focusing type CNLS equations with all $\s_l =1$, $l=1,2,...,N$. This study shows that the solutions 
 of mixed CNLS equations with  $N=2$ case can be generalized to arbitrary $N$ case just by allowing the number of components to run from $2$ to $N$ and redefining $\kappa_{ij}$'s suitably.
 
The procedure can be well understood by considering the example of writing down the soliton solutions of Eq. \eqref{1} for the case $N=3$. 
\subsubsection{One soliton solution}
The one soliton solution  of mixed 3-CNLS equations obtained by Hirota's method can be written as  
\begin{subequations}
\bear
\left(
\begin{array}{c}
q_1\\
q_2 \\
q_3
\end{array}
\right) 
&= &
\left(
\begin{array}{c}
\alpha_1^{(1)}\\
\alpha_1^{(2)} \\
\alpha_1^{(3)}
\end{array}
\right)\frac{e^{\eta_1}}{1+e^{\eta_1+\eta_1^*+R}}, \label{rc2} 
\eear
where 
\bear
\eta_1=k_1(t+ik_1z), \;\;
e^R =\frac{\kappa_{11}}{(k_1+k_1^*)},\;\; \label{rc3}
\eear
\ees
in which $ \kappa_{11} =\frac{\mu\left(\s_1|\al_1^{(1)}|^2+\s_2|\al_1^{(2)}|^2+\s_3|\al_1^{(3)}|^2\right)}{(k_1+k_1^*)}$  and without loss of generality we assume either $\s_1=1$, $\s_2=\s_3=-1$ or $\s_1=\s_2=1$, $\s_3=-1$. As in the case of $N=2$, Sec. III A, the solution is singular if $\s_1|\al_1^{(1)}|^2+\s_2|\al_1^{(2)}|^2+\s_3|\al_1^{(3)}|^2 \le 0$. Otherwise the solution is regular.
 It can be noticed that for any other combination of $\s_l$'s also the above solution satisfies Eq. \eqref{1}, for $N=3$.

\subsubsection{Two soliton solution}
The two soliton solution for the $N=3$ case is found to possess the same  form of Eq. \eqref{9}, with $j=1,2,3$, and $\kappa_{ij} $  is given by 
\bear
\kappa_{ij}= \frac{\mu\left(\sigma_1\alpha_i^{(1)}\alpha_j^{(1)*}+\sigma_2\alpha_i^{(2)}\alpha_j^{(2)*}+\sigma_3\alpha_i^{(3)}\alpha_j^{(3)*}\right)}
{\left(k_i+k_j^*\right)},\;i,j=1,2,
 \label{rc4}
\eear
where $\s_l$'s, $l=1,2,3$, can take the value either $+1$ or $-1$.
Here also the non-singular solution exists for the conditions \eqref{r2b}, \eqref{r2c}, and \eqref{r2f} with the redefinition of $\kappa_{ij}$'s as in Eq. \eqref{rc4}.

\subsubsection{Three and multisoliton solutions}
A similar analysis can be done for the multisoliton solutions  of the multicomponent case with arbitrary $N$.  Particularly  the three soliton solution of the mixed 3-CNLS equations , Eq. \eqref{1} with $N=3$, can be identified to have the form of three soliton solution for the $N=2$ case with $j$ running from  $1$ to $3$ (that is, now we have three components $q_1$, $q_2$, and $q_3$) and here $\kappa_{ij} $ is redefined  as 
\begin{figure}
	\centering
		\caption{Stationary singular three soliton solution  for $ N=3$ case.}
	\label{fig:fig8}
\end{figure}
\bear
\kappa_{ij}= \frac{\mu\left(\sigma_1\alpha_i^{(1)}\alpha_j^{(1)*}+\sigma_2\alpha_i^{(2)}\alpha_j^{(2)*}+\sigma_3\alpha_i^{(3)}\alpha_j^{(3)*}\right)}
{\left(k_i+k_j^*\right)},\;i,j=1,2,3, \label{rv}
\eear
where $\s_l$'s, $l=1,2,3,$ can take the value either $+1$ or $-1$ (see also Eq. (10) of Ref. \cite{Kanna03}). 

\begin{figure}
	\centering
	\caption{Shape changing (intensity redistribution)  collision of two solitons in the mixed CNLS system, for  $N=3$ case, exhibiting same kind of shape changes for a given soliton in all the three components.}
	\label{fig:fig9}
\end{figure}

It can also be noticed that the stationary singular solution for $N=3$ case given in Ref. \cite{kanna2004} results from the above mentioned three soliton solution for the choice 
\bear
\alpha_1^{(1)}=-e^{\eta_{10}},\; \alpha_2^{(2)}= e^{\eta_{20}},\;  
\alpha_3^{(3)}=-e^{\eta_{30}}, \; \alpha_i^{(j)}=0,\; k_{jI}=0, \mu=1,\; i \neq j,\; i,j=1,2,3,\label{ss2}
\eear
where  $\eta_{j0}$'s, $j=1,2,3$, are real parameters. The resulting limiting form reads in terms of hyperbolic functions as given in Appendix A. This singular solution at $z=0$ is shown in Figure \ref{fig:fig8}. The parameters are chosen as $\alpha_1^{(1)}=-1,\;$ $\alpha_2^{(2)}=1,\;  $
$\alpha_3^{(3)}=-1, \;$$ \alpha_i^{(j)}=0, i \neq j,\; i,j=1,2,3,$ $k_{1R}= 0.8$, $k_{2R}= 0.5$, $k_{3R}=0.4 $ , and $\mu=1$.

This procedure can be generalized further to obtain multisoliton solutions  of the multi-component case with arbitrary $N$.  For completeness we present the determinant form of the $N$-soliton solution of $N$-component case in Appendix B, following the lines of Ref. \cite{Ablowitz} for the Manakov case. 

\begin{figure}
	\centering
		\caption{Shape changing (intensity redistribution) collision of two solitons in the mixed CNLS system, for $ N=3$ case, exhibiting same kind of shape changes for a given soliton in the $q_1$ and $q_3$ components and an exactly opposite collision scenario  in the $q_2$ component.}
	\label{fig:fig10}
\end{figure}
\subsection{Collision scenario in multicomponent cases}
As we increase the number of components the collision behaviour becomes more interesting. For example, we consider the  collision of  two solitons in three component ($N=3$) mixed CNLS system.  We study the collision dynamics for the following two possible combinations of $\s$'s. For illustration, we present two nontrivial scenarios with two different choices of $\s_i$'s.

\noindent{\bf Case (i):}\;\;\underline{$\s_1= 1$, $\s_2=\s_3=-1$}\\
For this case, one possible parametric choice for non-singular solution is given by
 $k_1 = 1.0+i$, $k_2 = 0.9-i$, $\alpha_1^{(1)} =\alpha_2^{(1)}=1+i$, 
$\alpha_1^{(2)} = 0.2+0.4i$, $\alpha_2^{(2)} = 0.7+0.2i$,  $\alpha_1^{(3)} = 0.1+0.3i$,  $\alpha_2^{(3)} = 0.4+0.1i$, and $\mu=1$. We plot the two soliton solution corresponding to this parameter choice in Fig. \ref{fig:fig9}. The figure shows that after collision there is an enhancement (suppression)  of intensities (amplitudes) for a given soliton (say soliton $S_1$ ($S_2$)) in all the three components.
Here also one can verify that the difference of intensities is conserved according to the conservation law 
\bear
|A_1^{l\mp}|^2-|A_2^{l\mp}|^2-|A_3^{l\mp}|^2=\frac{1}{\mu}, \;\;l=1,2.
\eear 

\noindent {\bf Case (ii):}\;\; \underline{$\s_1= \s_2=1$ $\s_3=-1$}\\
Next we consider the above possible choice for $\s$'s. The nonsingular intensity plots of solitons $S_1$ and $S_2$ are shown in Fig. \ref{fig:fig10}. The parameters are chosen as 
$k_1 = 1.0+i$, $k_2 = 0.9-i$, $\alpha_1^{(1)} =1+i$, $\alpha_2^{(1)}=\frac{39-80i}{89}$, 
$\alpha_1^{(2)} = 0.2+0.4i$, $\alpha_2^{(2)} = 1$,  $\alpha_1^{(3)} = \frac{39+80i}{89}$,   $\alpha_2^{(3)} = 0.3+0.2i$ and $\mu=1$. This figure shows that after collision the intensity of soliton $S_1$($S_2$) in the first and third components gets enhanced (suppressed) while in the second component $S_1$($S_2$) is  suppressed (enhanced) in its intensity. This is a consequence of the conservation given by the relation 
\bear
|A_1^{l-}|^2 + |A_2^{l-}|^2- |A_3^{l-}|^2= |A_1^{l+}|^2 + |A_2^{l+}|^2- |A_3^{l+}|^2 =\frac{1}{\mu}, \;\;l=1,2.  \label{22}
 \eear

Thus for the two soliton solution of the $N$-component case the shape changing (intensity redistribution) collision occurs according to the relation 
\bear
 \sum_{l=1}^N \s_l|A_l^{j-}|^2 =  \sum_{l=1}^N \s_l|A_l^{j+}|^2 
 =\frac{1}{\mu} ,\;\; j=1,2. \label{23}
 \eear 
 However the elastic collision occurs for the choice
 \bear
\frac{\al_1^{(1)}}{\al_2^{(1)}}=\frac{\al_1^{(2)}}{\al_2^{(2)}}=...=\frac{\al_1^{(N)}}{\al_2^{(N)}}. \label{24}
\eear

One can also observe that multisoliton solutions for the case $N>2$ also undergo the above kind of shape changing (intensity redistribution) collisions but with more possible ways of energy exchange.

\section{Conclusion}
In this paper we have obtained the bright soliton type solutions of mixed CNLS Eq. \eqref{1} by applying Hirota's bilinear method. These solutions admit both singular and non-singular behaviours depending upon the choice of the soliton parameters. The condition for the existence of non-singular one and  two soliton  solutions for the  $N=2$ case are identified first. Analysing the corresponding collision behaviour reveals the fact that the solitons undergo fascinating shape changing (intensity redistribution) collisions with similar changes occurring in both components, which is not possible in the well known Manakov system.
This shape changing (intensity redistribution) collision occurs with a redistribution of intensities among the solitons, spread up in two components, in a particular fashion, where the intensity difference of the solitons between the two components is preserved after collision, and amplitude dependent phase-shifts as well as change in relative separation distances also occur. We have extended this study  to obtain multicomponent multisoliton solutions. Numerical plottings of the  solutions show that similar shape changing (intensity redistribution) collision behaviour are also observed for the multicomponent case with $N>2$ as in the case of $N=2$ but with many possible ways of shape variation. Still it is an open question to identify the regions in which system (1) admits  bright-dark, dark-bright, dark-dark  soliton solutions. Our study gives an adequate understanding of collision of bright-bright solitons arising in system \eqref{1} for mixed signs of nonlinearities. We believe that this kind of study will be of interest in the description of magnetic excitations over an anti-ferromagnetic vacuum, electromagnetic pulse propagation in left handed materials and so on.  In particular one of the most interesting properties of the bright solitons that we have identified in the present work is that the two components of a soliton can be simultaneously amplified during a collision process. Using this property, in principle it becomes possible to promote the collision process to the rank of a highly efficient amplification process without noise generation, in which the gain can be tuned over a relatively large range through a careful choice of pre-collision parameters. However, there still remains a lot of work to be done to make the fascinating concept of amplifiers with zero noise figure as practical device for optical communication systems. For example, an important and challenging issue will be to determine whether such amplification process can survive in the presence of strong perturbations or in the presence of propagation instabilities.

\section*{Acknowledgments}
T. K. acknowledges the Ministrie de l' Education Nationale, de la Recherche et de la Technology for offering a Research Associate fellowship. 
The work of M. L. is supported  by the Department of Science and Tecnology, Government of India, research project. N. A. acknowledges support from the Australian Research Council.

\appendix
\section{Singular stationary three soliton solution for  N=3 case}
In this appendix we present the  singular stationary three soliton solutions of mixed 3-CNLS equations. Considering the three soliton solution given by Eq. \eqref{18} but now the $\kappa_{ij}$'s are defined as in Eq. \eqref{rv}, the limiting form for the specific choice of parameters given by Eq. \eqref{ss2} can be deduced as
\bes
\bear
q_1 &=& \frac{-2k_{1R} \sqrt{\left(\frac{(k_{1R}+k_{2R})(k_{1R}+k_{3R})}{(k_{2R}-k_{1R})(k_{3R}-k_{1R})} \right)}
\left[\mbox{cosh}(A_1)+ \left|\frac{(k_{2R}+k_{3R})}{(k_{2R}-k_{3R})}\right|\mbox{cosh}(B_1)\right]e^{ik_{1R}^2z}}{D},\\
q_2 &=& \frac{2k_{2R} \sqrt{\left(\frac{(k_{1R}+k_{2R})(k_{2R}+k_{3R})}{(k_{2R}-k_{1R})(k_{3R}-k_{2R})} \right)}
\left[ \mbox{cosh}(A_2)-\left|\frac{(k_{1R}+k_{3R})}{(k_{1R}-k_{3R})}\right|\mbox{cosh}(B_2)\right]
e^{ik_{2R}^2z}}{D},\\
q_3 &=& \frac{2k_{3R} \sqrt{\left(\frac{(k_{1R}+k_{3R})(k_{2R}+k_{3R})}{(k_{3R}-k_{1R})(k_{3R}-k_{2R})} \right)}
\left[ \mbox{sinh}(A_3)+\left|\frac{(k_{1R}+k_{2R})}{(k_{2R}-k_{1R})}\right|\mbox{sinh}(B_3)\right]
e^{ik_{3R}^2z}}{D},
\eear
where
\bear
D &=& \mbox{cosh}(D_1)+\left|\frac{(k_{1R}+k_{2R})(k_{1R}+k_{3R})}{(k_{2R}-k_{1R})(k_{3R}-k_{1R})} \right|\mbox{cosh}(D_2) \nonumber\\
&&- \left|\frac{(k_{1R}+k_{2R})(k_{2R}+k_{3R})}{(k_{2R}-k_{1R})(k_{2R}-k_{3R})} \right|\mbox{cosh}(D_3)
 -\left|\frac{(k_{2R}+k_{3R})(k_{1R}+k_{3R})}{(k_{2R}-k_{3R})(k_{3R}-k_{1R})} \right|\mbox{cosh}(D_4), \\
A_1& =& (k_{2R}+k_{3R})t+\eta_{20}+\eta_{30} 
+\frac{1}{2}\mbox{ln}\left[\frac{(k_{2R}-k_{1R})(k_{3R}-k_{1R})(k_{3R}-k_{2R})^2}{16  k_{2R}^2k_{3R}^2(k_{1R}+k_{2R})(k_{1R}+k_{3R})(k_{2R}+k_{3R})^2}\right],\nonumber \\
B_1& =& (k_{2R}-k_{3R})t+\eta_{20}-\eta_{30}
+\frac{1}{2}\mbox{ln}\left[\frac{(k_{1R}-k_{2R})(k_{1R}+k_{3R})k_{3R}^2}
{ k_{2R}^2(k_{1R}+k_{2R})(k_{1R}-k_{3R})}\right],\nonumber \\
A_2& =& (k_{1R}+k_{3R})t+\eta_{10}+\eta_{30}
+\frac{1}{2}\mbox{ln}\left[\frac{(k_{2R}-k_{1R})(k_{3R}-k_{1R})^2(k_{3R}-k_{2R})}{16  k_{1R}^2k_{3R}^2(k_{1R}+k_{2R})(k_{1R}+k_{3R})^2(k_{2R}+k_{3R})}\right],\nonumber \\
B_2& =& (k_{1R}-k_{3R})t+\eta_{10}-\eta_{30}
+\frac{1}{2}\mbox{ln}\left[\frac{(k_{1R}-k_{2R})(k_{2R}+k_{3R})k_{3R}^2}{k_{1R}^2(k_{1R}+k_{2R})(k_{2R}-k_{3R})}\right],\nonumber \\
A_3& =& (k_{1R}+k_{2R})t+\eta_{10}+\eta_{20}
+\frac{1}{2}\mbox{ln}\left[\frac{(k_{3R}-k_{1R})(k_{2R}-k_{1R})^2(k_{3R}-k_{2R})}{16  k_{1R}^2k_{2R}^2(k_{1R}+k_{2R})^2(k_{1R}+k_{3R})(k_{2R}+k_{3R})} \right],\nonumber \\
B_3& =& (k_{1R}-k_{2R})t+\eta_{10}-\eta_{20}
+\frac{1}{2}\mbox{ln}\left[\frac{(k_{3R}-k_{1R})(k_{2R}+k_{3R})k_{2R}^2}
{ k_{1R}^2(k_{1R}+k_{3R})(k_{3R}-k_{2R})} \right],\nonumber \\
D_1& =& (k_{1R}+k_{2R}+k_{3R})t+\eta_{10}+\eta_{20}+\eta_{30} \nonumber \\
&&+\mbox{ln}\left[\frac{(k_{1R}-k_{2R})(k_{1R}-k_{3R})(k_{2R}-k_{3R})}{8  k_{1R}k_{2R}k_{3R}(k_{1R}+k_{2R})(k_{1R}+k_{3R})(k_{2R}+k_{3R})} \right],\nonumber \\
D_2 &=&  (k_{1R}-k_{2R}-k_{3R})t+\eta_{10}-\eta_{20}-\eta_{30}
+\mbox{ln}\left[\frac{2(k_{2R}+k_{3R})k_{2R}k_{3R}}{ k_{1R}
(k_{2R}-k_{3R})} \right],\nonumber \\
D_3 &=& (k_{1R}-k_{2R}+k_{3R})t+\eta_{10}-\eta_{20}+\eta_{30}
+\mbox{ln}\left[\frac{(k_{3R}-k_{1R})k_{2R}}{2 k_{1R} k_{3R}
(k_{1R}+k_{3R})}\right], \nonumber \\
D_4 &=& (k_{1R}+k_{2R}-k_{3R})t+\eta_{10}+\eta_{20}-\eta_{30}
+\mbox{ln}\left[\frac{(k_{2R}-k_{1R})k_{3R}}{2 k_{1R} k_{2R}
(k_{1R}+k_{2R})} \right].
\eear
\ees
Particularly, the stationary solution corresponding to the choice given in Eq. \eqref{ss2} can be easily checked to be the same as the  previously reported form given by Eq. (19) in Ref. \cite{kanna2004}. This clearly shows that the more general soliton solutions presented in this paper admit singular solutions as special cases which behave as regular and bounded solutions in specific regions.
\section{Multicomponent multisoliton solutions}
To write down the multicomponent multisoliton solutions in a formal way we  define the following $(1 \times N)$ row matrix  $C_s$ , $(N \times 1)$ column matrices  { $\psi_j$} , ${\phi}$,  $j,s = 1,2,...,N$, and the  $(N \times N)$ matrix ${\s}$:
\begin{subequations}
\label{app}
\bear
{C_s} = -\left(\alpha_1 ^{(s)}, \alpha_2 ^{(s)},..., \alpha_N^{(s)}\right),  \;\;
 \psi_j &= &\left(
\begin{array}{c}
\alpha_j ^{(1)}\\
 \alpha_j ^{(2)}\\
 \vdots \\
 \alpha_j ^{(N)}
 \end{array}
 \right),\;\;\;
 {\phi}=
 \left(\begin{array}{c}
e^{\eta_1}\\
 e^{\eta_2}\\
 \vdots \\
 e^{\eta_N} 
 \end{array}
 \right), \;\; j,s =1,2,...,N, \nonumber\\
 {\bf \s}= \left(
\begin{array}{cccc}
{\s_1} & {0} & ...&0\\
0 & {\s_2} & ...&0 \\
\vdots & \vdots & \ddots&\vdots  \\
0 & 0 & ...&{\s_N} 
\end{array}
\right),
 \eear
 where $\s_j$, $j=1,2,...,N$, can take value either $+1$ or $-1$.
Then the $N$-soliton solution of $N$-CNLS system \eqref{1} with mixed signs of nonlinearities can be written as 
\bear
q_s = \frac{g^{(s)}}{D},\;\;\; s=1,2,3,...,N,
\eear
where
\bear
g^{(s)}= \left|
\begin{array}{ccc}
{A} & {I} & {\phi}\\
 -{I} &{B} & 0 \\
0 & {C}_s & 0 
\end{array}
\right|,\;\;\;\;\;
D = 
\left|
\begin{array}{cc}
{A} & {I} \\
 -{I} & {B} 
 \end{array}
 \right|,
 \eear
in which  $s$ denotes the component. Here  ${I}$ is $(N \times N)$ unit matrix and the $(N \times N)$ matrices ${A}$ and ${B}$ are defined as
 \bear
{A}_{i,j} = \frac{e^{\eta_i+\eta_j^*}}{k_i+k_j^*}, \;\;\;
 {B}_{i,j} = \kappa_{ji}= \frac{\mu\left({\psi_i\dagger}{\bf\s}{\psi_j}\right)}{k_i^*+k_j},\;\;i,j=1,2,...,N,
 \eear
 \end{subequations}
  where $\eta_i = k_i(t+ik_iz)$, $k_i$ is complex, $\dagger$ represents the transpose conjugate. Here we remark that though presenting the solutions in determinant form seems to be compact, one has to explicitly write down the solutions as we have presented in Secs. II - V, for a complete characterization and analysis of the solution.  This way of expressing the solutions explicitly is also useful to identify the particular parameter choice  for which the singular stationary $N$-soliton solution of $N$-component case results from the general solutions. In particular, by generalizing the Eqs. \eqref{ss1} and \eqref{ss2} one can identify that the singular stationary $N$-soliton solution of the $N$-component case results from the above solution \eqref{app} for the choice $\al_i^{(i)}=(-1)^i e^{\eta_{i0}}$, $i=1,2,...,N$, and  $\al_i^{(j)} =0$,\;$k_{jI}=0$, $\mu=1$, where $i \neq j$, $i,j = 1,2,3,...,N$ and $e^{\eta_{i0}}$'s are arbitrary real parameters.

\newpage
\section*{Figure Captions}
Fig. 1: Singular one soliton solution of Eq. \eqref{1} for $ N=2$ case.

Fig. 2: Regular one soliton solution of Eq. \eqref{1} for $ N=2$ case.
	
Fig. 3: Stationary singular two soliton solution  for $ N=2$ case.

Fig. 4: Shape changing (intensity redistribution) collision of two solitons in the mixed CNLS system for $ N=2$ case.

Fig. 5: Elastic  collision of two solitons in the mixed CNLS system for  $N=2$ case.

Fig. 6: Shape changing (intensity redistribution) collision of three solitons in the mixed CNLS system for  $N=2$ case.

Fig. 7: Elastic collision of three solitons in the mixed CNLS system for the $N=2$ case.

Fig. 8: Stationary singular three soliton solution  for $ N=3$ case.

Fig. 9: Shape changing (intensity redistribution) collision of two solitons in the mixed CNLS system, for  $N=3$ case, exhibiting same kind of shape changes for a given soliton in all the three components.
	
Fig. 10: Shape changing (intensity redistribution) collision of two solitons in the mixed CNLS system, for $ N=3$ case, exhibiting same kind of shape changes for a given soliton in the $q_1$ and $q_3$ components and an exactly opposite collision scenario  in the $q_2$ component.
\end{document}